\documentstyle[11pt,paspconf,psfig]{article}

\begin{document}

\title{Molecular Absorption Lines at High Redshift}

\author{Tommy Wiklind}
\affil{Onsala Space Observatory, S--43992 Onsala, Sweden}

\author{Fran\c{c}oise Combes}
\affil{Observatoire de Paris, DEMIRM, 61 Av. de l'Observatoire\\
F--75014 Paris, France}

\begin{abstract}
The four known molecular absorption line systems at redshifts
z$=$0.25--0.89 are discussed. Two of the systems originate in
the galaxy hosting the `background' AGN, while in two cases
the absorption occurs in truely intervening galaxies. Indirect
evidence for the existence of very small scale structure in the
molecular ISM is presented. A new absorption line at z$=$0.89
towards PKS1830--211 opens up interesting possibilities to
accurately derive the mass of the lensing galaxy.
\end{abstract}


\section{Introduction}

Galaxy evolution in general and star formation in particular are
directly associated with molecular gas. The study of this dense and cold
component of the interstellar medium (ISM) in distant galaxies
gives important information about the conditions for star formation at
different epochs during the evolution of the universe.
In this respect, molecular absorption and emission lines are
complementary to each other. Whereas emission lines give information
about global properties, such as the total gas content and rotational
velocity, the absorption lines probe the molecular ISM in greater detail,
with more sensitivity and on spatial scales limited only by the angular
extent of the background source.
In addition, emission and absorption lines probe different types of
molecular gas. \\
\noindent
{\bf Emission.}
The observed property for molecular {\em emission} is the
velocity integrated line intensity, which for a galaxy at redshift
z can be expressed as $\int T_{\rm A}^{*} dV \approx
\left(\Omega_{\rm s}/\Omega_{\rm b}\right) T_{\rm b} \Delta V
\left(1+z\right)^{-1}$. The ratio of the solid angle of the source,
$\Omega_{\rm s}$, and the telescope beam, $\Omega_{\rm b}$, represents
the beam filling factor (valid as long as $\Omega_{\rm s} \leq \Omega_{\rm b}$).
$T_{\rm b}$ is the intrinsic brightness
temperature, where the observed intensity is diminished by
$\left(1+z\right)$ and $\Delta V$ is the width of the profile.
For optically thick emission, the brightness temperature is
$T_{\rm b} = \left[J(T_{\rm x})-J(T_{\rm bg})\right]$, where
$J(T) = \left(h\nu /k\right)/\{\exp (h\nu /k) - 1\}^{-1}$. $T_{\rm x}$
and $T_{\rm bg}$ are the excitation temperature and the temperature
of the cosmic background radiation, respectively.
For $T_{\rm x}$ somewhat higher than $T_{\rm bg}$, the brightness
temperature is approximately proportional to $T_{\rm x}$, while it
vanishes when $T_{\rm x} \rightarrow T_{\rm bg}$.
For a fixed--size galaxy, the observed integrated line intensity
$\int T_{\rm A}^{*} dV \propto \Omega_{\rm b}^{-1} \left(1+z\right)^{3}
D_{\rm L}^{-2} \Delta V T_{\rm x}$, where $D_{\rm L}$ is the luminosity
distance.\\
\noindent
{\bf Absorption.}
The observed property of a molecular {\em absorption} line is the
velocity integrated opacity. Assuming that the population of
rotational levels can be described by a single temperature $T_{\rm rot}=
T_{\rm x}$, not necessarily equal to the kinetic temperature, the
integated opacity can be expressed as:
$$
\int \tau_{\nu} dv \propto {N_{\rm tot}\over T_{\rm x}} \mu_{0}^{2}
\left(1-\exp{(-h\nu/kT_{\rm x})}\right)
\approx {N_{\rm tot} \over T_{\rm x}^{2}} \mu_{0}^{2}\ \ ,
$$
where $N_{\rm tot}$ is the total column density of the molecule in
question and $\mu_{0}$ the permanent dipole moment. This expression
is strictly speaking true only for linear molecules and transitions
from the ground state for which $h\nu << kT_{\rm x}$, but is similar
for more complex species and higher order transitions.

This different dependence on the excitation temperature for absorption
and emission lines illustrates how absorption measurements are
sensitive to excitationally cold gas, whereas emission lines probe
the warm gas. Under normal circumstances the excitation of molecular
rotational transitions is dominated by collisions. An excitationally
cold gas therefore corresponds to a diffuse gas component and warm
gas to dense gas.
The inverse quadratic dependence on $T_{\rm x}$ for absorption line
observations can have profound influences on the estimates of total
column densitites, which will be discussed below.
Furthermore, the observed integrated opacity does not include a distance
dependence for the absorption line measurements.
Instead the sensitivity only depends on the strength of the background
continuum source. This means that the same kind of data that can be obtained
in the Milky Way (e.g. Lucas \& Liszt 1994, 1996), in Cen A at a distance
of $\sim$4\,Mpc (e.g. Wiklind \& Combes 1997a) can be obtained towards
PKS1830--211 at a luminosity distance $\sim$4\,Gpc (Wiklind \& Combes 1996a,
1997c).

\begin{table}
\caption{Properties of molecular absorption line systems.} \label{tbl-1}
\begin{center} \scriptsize
\begin{tabular}{lcccccrc}
\tableline
Source & z$_{\rm a}$\tablenotemark{a} & z$_{\rm e}$\tablenotemark{b} &
$N_{\rm CO}$ & $N_{\rm H_2}$ & $N_{\rm HI}$ &
A$_{V}^{\prime}$\tablenotemark{c} & $N_{\rm HI}/N_{H_2}$ \\
 & & & cm$^{-2}$ & cm$^{-2}$ & cm$^{-2}$ & & \\
\tableline \\
Cen A         & 0.00184 & 0.0018 & $1.0 \times 10^{16}$ & $2.0 \times 10^{20}$ 
& $1 \times 10^{20}$ & 50 & 0.5 \\
PKS1413+357   & 0.24671 & 0.247  & $2.3 \times 10^{16}$ & $4.6 \times 10^{20}$
& $1.3 \times 10^{21}$ & 2.0 & 2.8 \\
B3\,1504+377A & 0.67335 & 0.673  & $6.0 \times 10^{16}$ & $1.2 \times 10^{21}$
& $2.4 \times 10^{21}$ & 5.0 & 2.0 \\
B3\,1504+377B & 0.67150 & 0.673   & $2.6 \times 10^{16}$ & $5.2 \times 10^{20}$
& $<7 \times 10^{20}$ & $<$2 & $<$1.4 \\
B\,0218+357   & 0.68466 & 0.94   & $2.0 \times 10^{19}$ & $4.0 \times 10^{23}$
& $4.0 \times 10^{20}$ & 850 & $1 \times 10^{-3}$ \\
PKS1830--211A & 0.88582 & ?      & $2.0 \times 10^{18}$ & $4.0 \times 10^{22}$
& $5.0 \times 10^{20}$ & 100 & $1 \times 10^{-2}$ \\
PKS1830--211B & 0.88489 & ?      & $1.0 \times 10^{16}$\tablenotemark{d} &
$2.0 \times 10^{20}$ & $1.0 \times 10^{21}$ & 1.8 & 5.0 \\
PKS1830--211C & 0.19267 & ?      & $<6 \times 10^{15}$                   &
$<1 \times 10^{20}$ & $2.5 \times 10^{20}$ & $<$0.2 & $>$2.5 \\
\tableline
\end{tabular}
\end{center}
\tablenotetext{a}{Redshift of absorption line}
\tablenotetext{b}{Redshift of background source}
\tablenotetext{c}{Extinction corrected for redshift using a Galactic extinction law}
\tablenotetext{d}{Estimated from the HCO$^{+}$ column density
of $1.3 \times 10^{13}$\,cm$^{-2}$}
\tablenotetext{e}{21cm HI data taken from Carilli et al. 1992,
1993, 1997a,b. A spin--temperature of 100\,K and a area covering
factor of 1 was assumed}

\end{table}

\begin{figure}
\psfig{figure=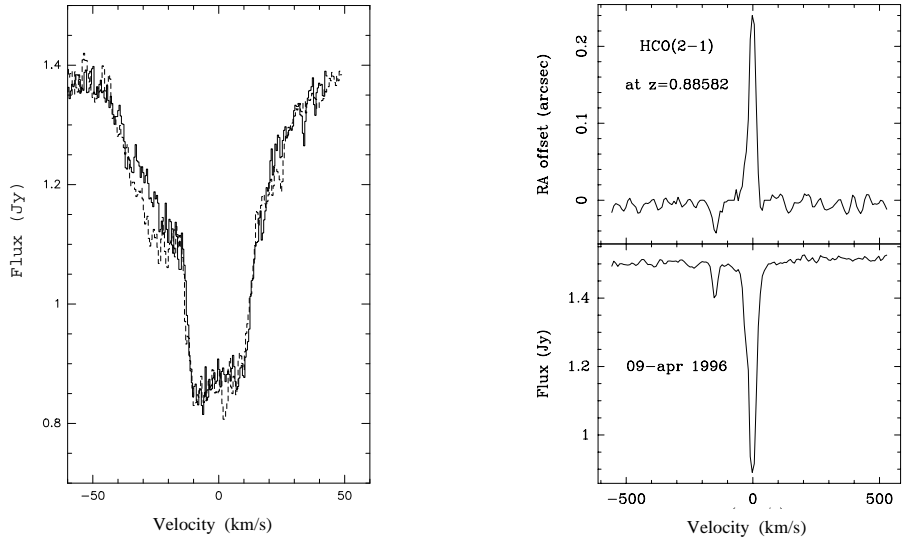,bbllx=20mm,bblly=100mm,bburx=185mm,bbury=200mm,width=12.0cm,angle=0}
\caption[]{HCO$^+$(2--1) absorption at z$=$0.88582 towards PKS1830--211
observed with the Plateau de Bure interferometer. {\bf a)} High resolution spectrum
at two different epochs (dashed/full drawn lines). {\bf b)} Low resolution spectra
(lower panel) showing two absorption line systems. The upper panel shows the shift
in RA of the phase center relative to that of the unobscured continuum flux.}
\end{figure}

\section{Known Molecular Absorption Line Systems}

There are four known molecular absorption line systems at high redshift:
z$=$0.25--0.89. These are listed in Table 1 together with data for the
low redshift absorption system seen toward the radio core of Cen A.
For the high redshift ones, a total of 15 different molecules have been
detected, in a total of 29 different transitions. This includes several
isotopic species: C$^{13}$O, C$^{18}$O, H$^{13}$CO, H$^{13}$CN.
As can be seen from Table 1, the inferred H$_2$ column densities varies
by $\sim 10^3$. The isotopic species are only detectable towards the
systems with the highest column densitites: B0218+357 and PKS1830--211.
The large dispersion in column densities is reflected in the large spread
in optical extinction $A_{\rm V}$ as well as the atomic to molecular ratio.
Systems with high extinction have 10--100 times higher molecular gas
fraction than those of low extinction. However, as discussed by Combes \&
Wiklind (these proceedings), the relation between HI and molecular gas
along sightlines to background QSOs can be ambiguous.

\subsubsection{Absorption in the host galaxy.}
Two of the four known molecular absorption line systems are situated within
the host galaxy to the `background' continuum source: PKS1413+135 and
B3\,1504+377. The latter exhibits two absorption line systems with similar
redshifts, z$=$0.67150 and 0.67335 (Wiklind \& Combes 1996b). The separation
in restframe velocity is 330\,km\,s$^{-1}$. This is the type of signature
one would expect from absorption occuring in a galaxy acting as a gravitational
lens. However, in this case, as well as for PKS1413+135, high angular
resolution VLBI images show no image multiplicity, despite impact parameters
less than 0.1'' (e.g. Perlman et al. 1996; Xu et al. 1995). The continuum
source must therefore be situated within or near the obscuring galaxy.
The small impact parameters make the latter situation highly unlikely.

\subsubsection{Absorption in gravitational lenses.}
The two absorption line systems with the highest column densities occur
in galaxies which are truely intervening and each acts as a gravitational
lens to the background source: B0218+357 and PKS1830--211. In these
two systems we have detected several isotopic species, showing that
the main lines (at least) are saturated. Nevertheless, the absorption
lines do not reach the zero level, indicating that only part of the
continuum source is covered by obscuring molecular gas, but that this
gas is optically thick (e.g. Combes \& Wiklind 1995, Wiklind \& Combes
1996a, 1997c).
The lensed images of B0218+357 and PKS1830--211 consist of two main
components. By comparing the depths of the saturated lines with fluxes
of the individual lensed components, as derived from long radio wavelength
interferometer observations, we identified the obscuration to only one
of two main lensed components (Wiklind \& Combes 1995; 1996a). This
has subsequently been verified through mm--wave interferometer data
(Menten \& Reid 1996; Wiklind \& Combes 1997c; Frye et al. 1997).

\begin{figure*}
\psfig{figure=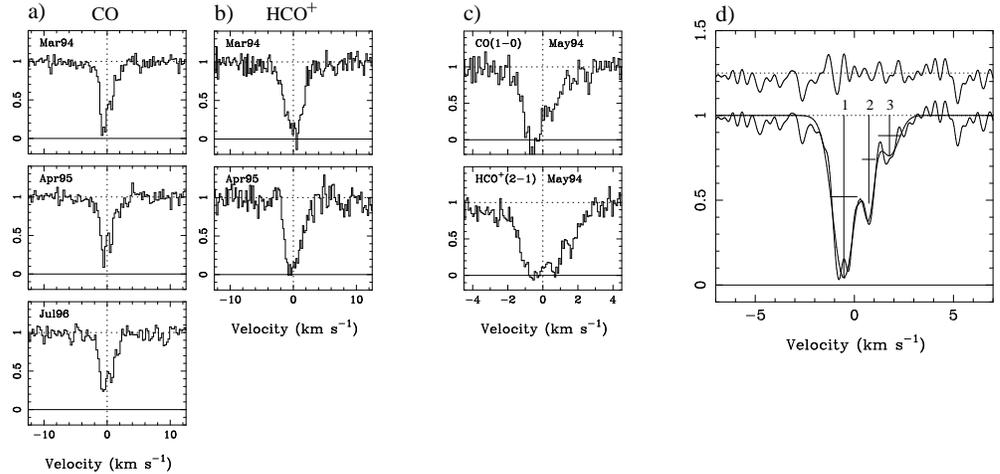,bbllx=15mm,bblly=142mm,bburx=186mm,bbury=227mm,width=13.2cm,angle=0}
\caption[]{Absorption lines towards PKS1413+135 at different epochs. {\bf a)} CO(1--0),
{\bf b)} HCO$^+$(2--1). In {\bf c)} the different line profiles of CO and HCO$^+$
are compared. {\bf d)} Decomposition of the CO(1--0) profile into 3 gauss components.}
\end{figure*}

\section{Small scale structure}

The angular resolution of the molecular absorption line observations
is set by the small, but finite, extent of the background continuum
source. The size of this region decreases with increasing frequency
and is located closer to the central engine than emission at
lower frequencies. VLBI observations at mm--wavelengths of the
flat--spectrum radio source 3C446 indicates a size $<$30\,$\mu$arcseconds
(Lerner et al. 1993), corresponding to less than 0.15\,pc.
The width of the absorption lines varies from less than 1\,km\,s$^{-1}$
(PKS1413+135) to $\sim$30\,km\,s$^{-1}$ (PKS1830--211).
The absorption profiles are likely to consist of a superposition of several
components (cf. Fig.\,1a). The widths implies that each sight line samples
molecular gas on parsec scales.
The observed absorption line is therefore an average over $\sim$0.1\,pc
in extent and $\sim$1\,pc in depth. If the molecular gas is structured
(density and/or temperature) on these scales, it can lead to observable
changes in the absorption profiles on relatively short time scales.

\begin{figure*}
\psfig{figure=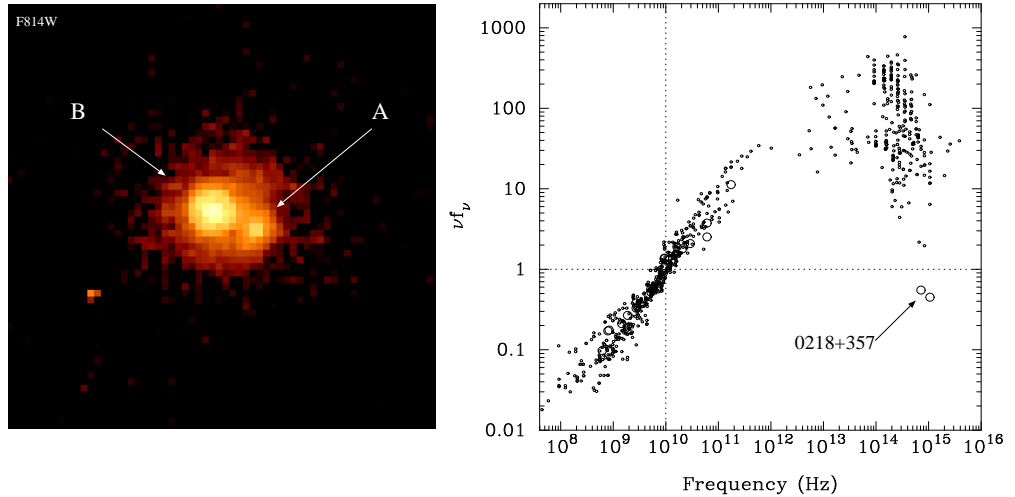,bbllx=19mm,bblly=156mm,bburx=188mm,bbury=243mm,width=13.2cm,angle=0}
\caption[]{{\bf Left.} HST F814W image of B0218+357. North is up, east to left.
{\bf Right.} Normalized SED for flat--spectrum radio QSOs. B0218+357 is marked by large
circles.}
\end{figure*}

\subsubsection{PKS1413+134:}
Variations have been seen in the CO(1--0) absorption line towards PKS1413+135
on a time scale of a few months (Wiklind \& Combes 1997b).
In Fig.\,2 we show the CO(1--0) and HCO$^+$(2--1) profiles from several
epochs. Whereas the HCO$^+$ profiles remain constant to within the noise,
the CO profiles exhibit changes in the depth with time. A close look
reveals that it is only one of three components in the absorption
profile that shows significant changes. By fitting three Gauss components
to the profiles (see Fig.\,2d), we found that the {\em ratio} of the
integrated opacities of component 1 and 2 has changed by a factor
$2.3 \pm 0.3$ between May 1994 and July 1996, in the sense that
component 1 had a higher opacity relative to component 2 in May 1994.
The continuum flux of PKS1413+135 decreased by a factor of 3 between
May 1994 and July 1996. This apparent correlation between variations
and the continuum flux suggests that the variations in opacity are
likely to be associated with structural changes in the AGN during an
outburst. The scale of such a structural change is unknown, but likely
to be very small. 

There are additional features in PKS1413+135 which suggest the presence
of small scale density structures in the molecular ISM.
The total column density observed towards PKS1413+135 is
$N_{\rm H} = 2 N_{\rm H_2}+N_{\rm HI} \approx 2 \times 10^{21}$\,cm$^{-2}$,
corresponding to an extinction
$A_{\rm V}^{\prime} \approx 2.0$\,mag\footnote{Corrected for the redshift
to correspond to $\sim A_{3000{\AA}}$ using a Galactic extinction law.}.
This is 25 times lower than the column inferred from the deficiency of low
energy X--ray photons, which implies $A_{\rm V} > 30$\,mag (Stocke et al. 1992).
Could this discrepancy be caused by an underestimate of the column density
derived from the absorption lines?

When deriving the column density it is necessary to estimate the rotation
temperature, i.e. the temperature governing the level population of a molecule.
If it is possible to observe two transitions of the same molecule, which is
usually the case, the ratio of the integrated opacities allows a derivation
of the excitation temperature for these two levels.
With the assumption of LTE conditions this temperature is equal to the 
rotation temperature. If an additional transition of the same molecule can
be observed, the set of two opacity ratios should give the same excitation
temperature. Towards PKS1413+135 we have observed the J$=$1--0,
2--1 and 3--2 transitions of HCO$^+$. Whereas the J$=$2--1/1--0 ratio
gives $T_{\rm x} = 5.4 \pm 0.7$ K, the J$=$2--1/3--2 ratio gives
$T_{\rm x} = 8.7 \pm 0.2$ K
Although this discrepancy suggests that the assumption of LTE conditions
does not hold, it is also compatible with LTE conditions and a multi--component
gas along the line of sight, where the different components are characterized
by different excitation temperatures (Wiklind \& Combes 1997c).
For instance, expressing the observed opacity as the sum of two components,
characterized by two excitation temperatures $T_1 < T_2$:
$$
\int \tau_{\nu} dV \propto N_{\rm tot} \left({\gamma \over f\left(T_1\right)}
+ {1-\gamma \over f\left(T_2\right)}\right)\ \ ,
$$
where $N_{\rm tot}$ is the total column density along the line of sight,
$f$ is a function of the excitation temperature (e.g. Wiklind \& Combes 1997b)
and $\gamma$ is the fraction of `cold' gas (i.e. $T_1$), a solution for
the discrepant $T_{\rm x}$ of HCO$^+$ in PKS1413+135 is compatible with
two gas components characterisized by $T_1 = 4$\,K and $T_2 = 18$\,K,
where $\sim$20\% of the total column density originates in the cold component
and the remaining $\sim$80\% in the warm component. The total column density
can be $\sim$10 times higher than the one estimated assuming a one component
gas, making it more compatible with the column estimated from the deficiency
of soft X--ray photons.
Since the excitation of molecular gas is dominated by collisions between the
molecule in question and H$_2$, a low excitation temperature corresponds to
diffuse gas, while a high excitation temperature corresponds to dense gas.
This means that a two component gas as described above corresponds to a
gas consisting of a dense and a diffuse part, implying structures with
a large density contrast on the scales probed by the line of sight, i.e.
of the order $\sim 10^{-2}$\,pc$^{3}$.

\subsubsection{B0218+357}
The molecular gas seen towards B0218+357 covers only one of the two lensed
images of the background source (Wiklind \& Combes 1995; Menten \& Reid
1996). Saturated lines of both $^{13}$CO and C$^{18}$O  have been
detected, while the C$^{17}$O transition remains undetected (Combes
\& Wiklind 1995). This gives a lower limit to the CO column density
which transforms to $N_{\rm H_2} \approx 4 \times 10^{23}$\,cm$^{-2}$
and an $A_{\rm V}^{\prime} \approx 850$\,mag. The absorption occurs in front
of the A--component, which is then expected to be completely invisible at
optical wavelengths. Nevertheless, analysis of archival HST data obtained
with the WFPC2 in broad V-- and I--band, show both components (Fig\,2a).
While the intensity ratio A/B of the two lensed images is 3.6 at radio
wavelengths (c.f. Patnaik et al. 1996), A/B$\approx$0.15
at optical wavelengths. The V$-$I values show no difference in reddening
for the A-- and B--component. Hence, there are no indications of excess
extinction in front of the A--component.
The B--component is situated close to the center of the lensing galaxy and
the optical flux from B--component is the sum of the lensed image and the
nucleus of the lensing galaxy.
Is the A--component an unobscured view of the background QSO, with the
molecular absorption occuring elsewhere? To answer this question we compared
the optical flux of the A--component (multiplied by a factor 1.3 to compensate
for the B--component using the magnification ratio of 3.6) with that of
similar flat--spectrum radio sources. Fluxes where taken from the literature,
corrected for the redshift of the source and then normalized at the restframe
frequency of 10\,GHz. In Fig.\,2b we plot the observed luminosities (normalized
$\nu f_{\nu}$ units). Despite a very large dispersion of optical luminosities
for the flat--spectrum sources, the optical luminosity of B0218+357, as derived from
the measured flux of the A--component, is clearly sub--luminous.
This implies that, unless B0218+357 is a very peculiar source, a significant
part of the optical emission region in component A is totally obscured,
with $A_{\rm V}$ being very high, while a small part
of it remains unobscured. Since the extent of the optical emission region
is very small, this suggests the presence of very small scale structure with
large density contrasts in the molecular ISM of the lensing galaxy.

\section{The complex absorption line system towards PKS1830--211}

The radio loud QSO PKS1830--211 is lensed by an intervening galaxy
and consists of two main images, designated the NE and SW images.
The lensing galaxy was detected through absorption of molecular
rotational lines, at a redshift z$=$0.88582 (Wiklind \& Combes 1996a).
The absorption only occurs in front of the SW component. This was
inferred from the depth of saturated lines and has been confirmed
through mm--wave interferometric observations (Wiklind \& Combes 1997c;
Frye et al. 1997). A second absorbing system (designated as C in Table
1) has been identified as a 21cm HI absorption occuring at z$=$0.19
(Lovell et al. 1996). This latter system is not seen in either molecular
absorption nor emission (Wiklind \& Combes 1997c). This HI absorption only
occurs in front of the NE component.

A second molecular absorption system has been detected at z$=$0.88489 (Wiklind
\& Combes 1997c). The new absorption line is separated by 147\,km\,s$^{-1}$
from the previously detected absorption (Fig.\,1b) and occurs towards the NE
component, in contrast to the main line which obscures the SW component.
This is clearly seen in Fig.\,1b, which shows the HCO$^+$(2--1) absorption
observed with the IRAM Plateau de Bure interferometer. The spectrum is
centered on the main line. The top panel shows the shift of the phase--center
along the $\alpha$--coordinate as a function of velocity. At velocities where
the continuum of the SW component is obscured by the main HCO$^+$ absorption,
the phase--center shifts towards the NE component (positive $\alpha$), while
the opposite is true for velocities corresponding to the weaker absorption.
Hence, the HCO$^+$ absorption probes two sight lines through the intervening
galaxy. Since the center of the lensing potential must be situated between
the two lensed images, the velocity separation of the two absorption lines
measures the rotational velocity of the gas. In order to interpret the observed
velocity separation in terms of rotation, it is necessary to know where in
the galaxy the sight lines occur. This can be done using existing lens
models (cf. Wiklind \& Combes 1997c), or through direct imaging of the
lensing galaxy (see Frye, these proceedings).
The second absorption system at z$=$0.885 has also been seen in the line
of HCN(1--0) and in 21cm HI absorption (Carilli et al. 1997b). The atomic
fraction is higher in the NE component (weaker molecular absorption) than 
in the SW component, in line with the apparent correlation previously
mentioned between the atomic and molecular fraction and total extinction
(See Table\,1).


\begin{references}
\reference Carilli, C.L., Perlman E.S., Stocke J.T. 1992, ApJ 400, L13 
\reference Carilli, C.L., Rupen, M.P., Yanny, B. 1993, ApJ 412, L59 
\reference Carilli, C.L., Menten, K.M., Reid, M.J., Rupen M.P.: 1997a,
ApJ 474, L89
\reference Carilli, C.L., Menten, K.M., Reid, M.J., Rupen, M.P.,
Claussen, M. 1997b, 13$^{th}$ IAP Colloqium: Structure and Evolution of
the IGM from QSO Absorption Line Systems, eds. P. Petitjean, S. Charlot
\reference Combes, F., Wiklind T. 1995, A\&A 303, L61
\reference Combes F., Wiklind T., 1996, in Cold Gas at High Redshift,
eds. M.N. Bremer, P. van der Werf, H.J.A. R\"{o}ttgering, C.L. Carilli,
Kluwer Academic Pub., p.\,215
\reference Combes, F., Wiklind T. 1997, ApJ 486, L59
\reference Frye B., Welch W. J., Broadhurst T. 1997, ApJ 478, L25
\reference Lerner M., B{\aa}th L., Inoue M. et al. 1993, A\&A 280, 117
\reference Lovell, J.E.J., Reynolds, J.E., Jauncey D.L., et al. 1996,
ApJ 472, L5
\reference Lucas R., Liszt H.S.: 1994, A\&A 282, L5
\reference Lucas R., Liszt H.S.: 1996, A\&A 307, 237
\reference Menten K.M., Reid M.J.: 1996, ApJ 465, L99
\reference Perlman E.S., Carilli C.L., Stocke J.T., Conway, J.S. 1996,
AJ 111, 1839
\reference Wiklind, T., Combes, F. 1994, A\&A 286, L9  
\reference Wiklind, T., Combes, F. 1995, A\&A 299, 382 
\reference Wiklind, T., Combes, F. 1996a, Nature, 379, 139 
\reference Wiklind, T., Combes, F. 1996b, A\&A 315, 86 
\reference Wiklind, T., Combes, F. 1997a, A\&A 324, 51 
\reference Wiklind, T., Combes, F. 1997b, A\&A 328, 48 
\reference Wiklind, T., Combes, F. 1997c, ApJ in press (astro--ph/9709141)
\reference Xu W., Readhead A.C.S., Pearson T.J., Polatidis A.G., Wilkinson P.N.
1995, ApJS 99, 297
\end{references}
\end{document}